\documentclass[11pt]{article}

\usepackage{a4wide}
\usepackage{xspace}

\usepackage[all]{xy}
\usepackage{amssymb}
\usepackage{amsmath}
\usepackage{graphics}
\usepackage{wrapfig}
\newtheorem{theorem}{Theorem}[section]

\newtheorem{proposition}[theorem]{Proposition}

\newtheorem{definition}[theorem]{Definition}

\newcommand{\push}{\textsc{push}\xspace}
\newcommand{\pop}{\textsc{pop}\xspace}
\newcommand{\pda}{\text{PDA}\xspace}
\newcommand{\dpda}{\text{DPDA}\xspace}

\newcommand{\act}{\mbox{${\cal A}${\sl ct}}}

\newcommand{\head}[1]{\mathit{head}{(#1)}}
\newcommand{\tail}[1]{\mathit{tail}{(#1)}}

\newcommand{\goes}[1]{\ensuremath{\stackrel{#1}{\longrightarrow}}}

\title{Note on Undecidability of Bisimilarity for \\ Second-Order
Pushdown Processes\\[5mm]}

\author{Petr Jan\v{c}ar \\[3mm] 
{\normalsize Department of Computer Science, FEI V\v{S}B-TU}\\
{\normalsize Technical University of Ostrava}\\
{\normalsize Czech Republic}\\
\texttt{petr.jancar@vsb.cz}
\and
Ji\v{r}\'{i} Srba \\[3mm] 
{\normalsize Department of Computer Science}\\
{\normalsize Aalborg University}\\ 
{\normalsize Denmark}\\
\texttt{srba@cs.aau.dk}}

\date{ }
\begin{document}

\maketitle

\vspace{-6mm}
\begin{abstract}
\noindent
Broadbent and G\"oller (FSTTCS 2012)
proved the undecidability of bisimulation equivalence for
processes generated by
$\varepsilon$-free second-order pushdown automata.
We add a few remarks concerning the
used proof technique, called Defender's forcing, 
and the related undecidability proof for  
first-order pushdown automata with $\varepsilon$-transitions 
(Jan\v{c}ar and Srba, JACM 2008).
\end{abstract}\vspace{3mm}

\noindent
Language equivalence of pushdown automata (\pda)
is a well-known problem in computer science community.
There are standard textbook proofs showing
the undecidability even for \emph{$\varepsilon$-free}
\pda, i.e. for \pda that have no $\varepsilon$-transitions;
such \pda are sometimes called \emph{real-time} \pda.
The decidability question of language equivalence 
for \emph{deterministic} \pda (\dpda)
was a famous
long-standing open problem. It was positively answered by 
Oyamaguchi~\cite{DBLP:journals/jacm/Oyamaguchi87}
for $\varepsilon$-free \dpda and later by 
S\'enizergues~\cite{Senizergues:TCS2001} for the whole class 
of \dpda.

Besides their role of language acceptors, 
\pda can be also viewed as generators of (infinite) labelled transition
systems; in this context it is natural to study another fundamental
equivalence, namely \emph{bisimulation equivalence}, also called
\emph{bisimilarity}. This equivalence
is finer than language equivalence, but the 
two equivalences in principle coincide on deterministic systems.

S\'enizergues~\cite{Senizergues:SIAM:05} 
showed an involved proof of the
decidability of bisimilarity for (nondeterministic)
$\varepsilon$-free \pda but also for \pda in which 
$\varepsilon$-transitions are deterministic, popping and 
do not collide with ordinary input $a$-transitions.
It turned out that 
a small relaxation, allowing for nondeterministic
popping $\varepsilon$-transitions, already leads to 
undecidability~\cite{DBLP:journals/jacm/JancarS08}.
In~\cite{DBLP:journals/jacm/JancarS08}, the authors of this note also
explicitly describe
a general proof technique called \emph{Defender's forcing}. 
It is a simple, yet powerful, idea related to the bisimulation game
played between Attacker and Defender; it was used, sometimes
implicitly, also in context of other hardness results
for bisimilarity on various classes of infinite state systems.

The classical PDA, to which we have been referring so far, are
the \emph{first-order} \pda in the hierarchy of 
\emph{higher-order} \pda that were introduced 
in connection with higher-order recursion
schemes already in 1970s.
The decidability question for equivalence of deterministic
$n$th-order \pda,
where $n\geq 2$,
seems to be open so far. A step towards a solution was made by
Stirling~\cite{DBLP:conf/concur/Stirling06} who 
showed
the decidability for a subclass of 
$\varepsilon$-free
second-order \dpda.

Recently 
Broadbent and G{\"o}ller~\cite{DBLP:conf/fsttcs/BroadbentG12} 
noted that the results in~\cite{DBLP:journals/jacm/JancarS08}, or
anywhere else in the literature, do not answer the decidability question for
bisimilarity of \emph{$\varepsilon$-free} second-order \pda.
They used the above mentioned technique of Defender's
forcing to show that this problem is also undecidable.
This result helps to further 
clarify the (un)decidability border, now in another
direction:
a mild use of second-order operations 
(on a stack of stacks)
is sufficient 
to establish undecidability without using
$\varepsilon$-transitions (that are necessary in the first-order
undecidability proof~\cite{DBLP:journals/jacm/JancarS08}).

The authors of~\cite{DBLP:conf/fsttcs/BroadbentG12} concentrate on
giving a complete self-contained technical construction yielding the
undecidability proof, however, they do not discuss in detail its relation to the
constructions in~\cite{DBLP:journals/jacm/JancarS08}. 
Here, in Section~\ref{sec:undecepsilon},
we try to concisely  present the idea of the relevant first-order proof 
from~\cite{DBLP:journals/jacm/JancarS08}, 
and then, in Section~\ref{sec:undecsecondorder}, we highlight the
idea in~\cite{DBLP:conf/fsttcs/BroadbentG12} that makes it 
possible to replace
the use of $\varepsilon$-transitions 
in the undecidability proof
with second-order operations.

We hope that this note may help to popularize 
the Defender's forcing technique,
and that it might be found useful by other researchers
tackling further open problems in the area. 

\section{Definitions}\label{sec:definitions}

A \emph{labelled transition system} (LTS) is a 
(possibly infinite) directed multigraph with action-labelled
edges. By a triple $s\goes{a}s'$, called a \emph{transition}, 
or an \emph{$a$-transition},
we denote
that there is 
an edge from node $s$ to node $s'$ labelled with $a$\,;
we also refer to the nodes as
to the \emph{states}.
A \emph{symmetric} binary relation $R$ on the set of states
is a \emph{bisimulation} if for any $(s,t) \in R$ and any 
transition $s \goes{a} s'$ there is a transition $t \goes{a} t'$ 
(with the same label $a$)
such that $(s',t') \in R$. Two states $s$ and $t$ are
\emph{bisimilar}, written $s \sim t$, if there is a bisimulation
containing $(s,t)$.

Bisimilarity is often presented
in terms of a two-player game
between Attacker (he) and Defender (she).
In the current game position, that is a pair of states $(s_1,s_2)$ 
in an LTS,
Attacker chooses a transition $s_j\goes{a}s'_j$ (for $j\in\{1,2\}$)
and Defender then chooses a transition
$s_{3-j}\goes{a}s'_{3-j}$; 
the pair $(s'_1,s'_2)$ becomes the new current position.
If one player gets stuck then the other player wins;
an infinite play is a win of Defender. It is easy to verify
that $s, t$ are
bisimilar iff Defender has a winning strategy when starting from
the position $(s,t)$.

An \emph{$\varepsilon$-free second-order pushdown system}
is a tuple $(Q,\Gamma,\act,\Delta)$ consisting of
four finite nonempty sets:
$Q$ contains the
\emph{control states}, 
$\Gamma$ the \emph{stack symbols},
$\act$ the
\emph{actions} (corresponding to classical input letters),
and $\Delta$ the \emph{rules} of the following three types:
\begin{equation}\label{eq:rules}
pX \goes{a} q \alpha, 
\  pX \goes{a} (q,\push),
\ pX \goes{a} (q, \pop),
\end{equation}
where  $p,q \in Q$, 
$X \in \Gamma$,
$a \in \act$, and $\alpha \in \Gamma^*$.
The LTS generated by $(Q,\Gamma,\act,\Delta)$
has the set $Q\times(\Gamma^+)^*$ as the set of states; 
a state is written in the form $q[\delta_1][\delta_2]\cdots [\delta_n]$
where $q$ is a control state and 
$\delta_i$ is a nonempty sequence of stack symbols (for $i=1,2,\dots,n$).
By $\varepsilon$ we denote the empty sequence; hence
$[\delta_1][\delta_2]\cdots [\delta_n]=\varepsilon$ when $n=0$.
The transitions in the generated LTS are induced by the rules from $\Delta$ as
follows:
\begin{itemize}
\item
the rule $pX\goes{a}q\alpha$ implies 
$p[X\gamma][\delta_2][\delta_3]\dots [\delta_n]\goes{a}
q[\alpha\gamma][\delta_2][\delta_3]\dots [\delta_n]$ if
$\alpha\gamma\neq\varepsilon$,
\\
and  $p[X\gamma][\delta_2][\delta_3]\dots [\delta_n]\goes{a}
q[\delta_2][\delta_3]\dots [\delta_n]$ if $\alpha\gamma=\varepsilon$\,;
\item the rule
$pX\goes{a}(q,\push)$ implies 
$p[X\gamma][\delta_2][\delta_3]\dots [\delta_n]\goes{a}
q[X\gamma][X\gamma][\delta_2][\delta_3]\dots [\delta_n]$\,;
\item the rule
$pX\goes{a} (q,\pop)$ implies 
$p[X\gamma][\delta_2][\delta_3]\dots [\delta_n]\goes{a}
q[\delta_2][\delta_3]\dots [\delta_n]$\,.
\end{itemize}
We remark that the definitions of second-order pushdown systems in the
literature vary in details that are insignificant for us.
If we restrict the rules to the type $pX\goes{a}q\alpha$ then we get 
\emph{$\varepsilon$-free first-order pushdown systems}.
In this paper we do not introduce $\varepsilon$-rules 
(of the types~(\ref{eq:rules}) with $a=\varepsilon$);
their restricted use in our paper is handled by a remark at the
respective place.

\section{Undecidability of bisimilarity for \pda with
$\varepsilon$-transitions}\label{sec:undecepsilon}

In this section, we briefly explain a result
from~\cite{DBLP:journals/jacm/JancarS08}, namely the undecidability of
bisimilarity for (normal, i.e. first-order) pushdown systems with 
popping $\varepsilon$-rules (of the type $pX\goes{\varepsilon}q$). 
The text closely follows the beginning of Section 5.1
from~\cite{DBLP:journals/jacm/JancarS08}, though it is 
a bit modified, concentrating on illustrating the ideas.

The undecidability result is 
achieved by a reduction from the following variant of
Post's Correspondence Problem (PCP). 
As usual, by a \emph{word} $u$ \emph{over an alphabet} we
mean a finite sequence of letters; $|u|$ denotes the length of $u$.

\begin{definition}
A \emph{PCP-instance} INST
is a nonempty sequence
$(u_1,v_1), (u_2,v_2), \dots, (u_n,v_n)$ of pairs of nonempty words
over the alphabet $\{A,B\}$ where
$|u_i|\leq |v_i|$ for all $i\in\{1,2,\ldots,n\}$.
An \emph{infinite initial solution} of INST, a \emph{solution} of INST for short,
is an infinite sequence of
indices $i_1, i_2, i_3, \ldots$ from the set $\{1,2,\ldots, n\}$
such that $i_1 {=} 1$ and the infinite words
$u_{i_1}u_{i_2}u_{i_3}\cdots$ and
$v_{i_1}v_{i_2}v_{i_3}\cdots$
are equal. A finite sequence $i_1, i_2, \dots, i_{\ell}$ is a
\emph{partial solution} of INST if $i_1 {=} 1$ and
$u_{i_1}u_{i_2}\cdots u_{i_\ell}$
is a prefix of $v_{i_1}v_{i_2}\cdots v_{i_\ell}$.
\\
The problem \emph{inf-PCP} asks if there is 
a solution
for a given INST.
\end{definition}

The next proposition can be shown by standard arguments, related to
simulations of \emph{nonterminating} Turing machine computations;
the respective reduction easily guarantees our 
technical condition $|u_i|\leq |v_i|$ 
(see also \cite{DBLP:journals/jacm/JancarS08}). 

\begin{proposition}
\label{prop:infandrecPCP}
Problem inf-PCP is undecidable; more precisely, 
inf-PCP is $\Pi^0_1$-complete.
\end{proposition}

We now consider
a fixed instance INST of inf-PCP,
i.e. $(u_1,v_1), (u_2,v_2), \dots, (u_n,v_n)$
as above.
Let us imagine the following game, played between Attacker (he)
and  Defender (she); this game is more abstract, it will be only later
implemented as the bisimulation game.

Starting with the one-element sequence
$i_1$, where $i_1=1$, Attacker repeatedly asks Defender to prolong 
the current sequence 
${i_\ell} {i_{\ell-1}}\dots {i_{1}}$ by one
$i_{\ell+1}\in\{1,2,\dots,n\}$ (of her choice),
to get ${i_{\ell+1}} {i_{\ell}}\dots {i_{1}}$. 
(We use prolongations to the left, to ease the later implementation by
a pushdown system.)
Attacker can thus ask Defender indefinitely, 
in which case the play is
a win for Defender, or he can
eventually decide to
switch to checking whether the current sequence
represents a partial solution, i.e., 
whether $u_{i_1} u_{i_2}\dots u_{i_{\ell}}$ is a prefix of
$v_{i_1} v_{i_2}\dots v_{i_{\ell}}$; the negative case is a win for
Attacker, the positive case is a win for Defender. 
In another formulation, 
the checking phase finds out
whether
 $(u_{i_\ell})^R (u_{i_{\ell-1}})^R\dots (u_{i_{1}})^R$ is equal to 
 a suffix
of 
 $(v_{i_\ell})^R (v_{i_{\ell-1}})^R\dots (v_{i_{1}})^R$, where 
 $w^R$ denotes the reverse of $w$.
It is obvious that 
\begin{equation}\label{eq:soliffwin}
\textnormal{INST has a solution
iff Defender has a winning strategy.} 
\end{equation}

With an eye to the later implementation of the game by pushdown rules,
we formulate an intermediate version of the game as follows.
(In fact, this intermediate game replaces the arguments 
given in~\cite{DBLP:journals/jacm/JancarS08} to justify
the rules of Fig.~\ref{fig:rulesystem}.)

\begin{itemize}
\item
(\emph{Generating phase}) 
\\
The game starts with a pair $(q_0\, i_1, q'_0\, i_1)$ where $i_1=1$
and $q_0, q'_0$ are auxiliary symbols that we can call ``control
states''.
Attacker repeatedly asks Defender to prolong 
both sequences in the current pair 
$(q_0\, {i_\ell}{i_{\ell-1}}\dots {i_{1}},q'_0\, {i_\ell} {i_{\ell-1}}\dots {i_{1}})$ 
by some $i_{\ell+1}\in\{1,2,\dots,n\}$,
thus creating the next current pair 
$(q_0\, {i_{\ell+1}} {i_{\ell}}\dots {i_{1}},q'_0\, {i_{\ell+1}}
{i_{\ell}}\dots {i_{1}})$. 
\item
(\emph{Switching phase}) 
\\
For any current pair 
\begin{equation}\label{eq:switchstart}
(q_0\, {i_\ell} {i_{\ell-1}}\dots {i_{1}},q'_0\, {i_\ell} {i_{\ell-1}}\dots {i_{1}})
\end{equation}
Attacker can decide to switch (to the verification): 
the control state in the left-hand sequence changes to $q_u$;
in the right-hand side sequence 
the control state changes to $q_v$ but before that
Defender can erase a chosen prefix 
${i_{\ell}}{i_{\ell-1}}\dots {i_{\ell-k}}$ and replace 
${i_{\ell-k-1}}$ with a suffix $w$ of $(v_{\ell-k-1})^R$; we thus get
\begin{equation}\label{eq:switchend}
(q_u\, {i_\ell} {i_{\ell-1}}\dots {i_{1}},q_v\, w\,{i_m} {i_{m-1}}\dots {i_{1}})
\textnormal{ where } m<\ell \textnormal{ and } 
w \textnormal{ is a suffix of } v_{i_{m+1}}\,. 
\end{equation}
\item
(\emph{Verification phase})
\\
Here the play is completely determined, verifying (step by step)
that 
 $(u_{i_\ell})^R (u_{i_{\ell-1}})^R\dots (u_{i_{1}})^R$ is equal
to
 $w\, (v_{i_m})^R (v_{i_{m-1}})^R\dots (v_{i_{1}})^R$.
The control state $q_u$ signals that $i_j$ is interpreted as 
$(u_{i_j})^R$, and $q_v$ signals that $i_j$ is interpreted as 
$(v_{i_j})^R$.
If a mismatch is encountered then Attacker wins, otherwise Defender wins.
\end{itemize}
Property~(\ref{eq:soliffwin}) obviously holds  
for the above (intermediate) game as well.
We now show that 
this game is implemented as the bisimulation game 
in the LTS generated by the pushdown system 
in Fig.~\ref{fig:rulesystem}, 
starting in the position $(q_0 I_1\bot, q'_0 I_1\bot)$.
We use the symbol $I_i$ instead of $i$; 
the ``bottom-of-the-stack'' symbol
$\bot$ is used for technical reasons.

\begin{figure}[ht]

\begin{tabular}{lllll}
(G1) rules: \hspace{1mm} & $q_0 \goes{g} t$ \\  
\vspace{2mm}
&\fbox{$q_0 \goes{g} p_i$} \hspace{5mm} & $q'_0 \goes{g} p_i$  
\\
&$t \goes{a_i} q_0 I_i$ &
$p_i \goes{a_i} q'_0 I_i$ \\
&& \fbox{$p_i \goes{a_j} q_{0} I_j$} & where $i\neq j$
\end{tabular}
\begin{tabular}{llll}
(S1) rules: \hspace{1mm} & $q_0 \goes{s} q_u$\\
&\fbox{$q_0(I^*)I_i \goes{s} q_vw$} \hspace{3mm} &
$q'_0(I^*)I_i \goes{s} q_vw$ \hspace{-2mm} &
for all suffixes $w$ of $v_i^R$\\
\end{tabular}

\medskip

\begin{tabular}{llll}
(V1) rules: \hspace{1mm} & 
$q_u I_i \goes{h(u_i^R)} \ q_u\,\, \tail{u_i^R}$ 
 \hspace{4mm} & $q_v I_i \goes{h(v_i^R)} \ q_v\,\, \tail{v_i^R}$ 
\\
& $q_u A \goes{a} q_u$ &  $q_v A \goes{a} q_v$ & \hspace*{2cm} \\
&$q_u B \goes{b} q_u$ &  $q_v B \goes{b} q_v$ \\
\end{tabular} 

\begin{quote}
\emph{Notation.} 
A rule $p\goes{a}q\alpha$ replaces the family 
$pX\goes{a}q\alpha X$ 
for all stack symbols $X$.
By $\head{w}$ we denote the first symbol of $w$;
$\tail{w}$ is the rest of $w$, and thus 
$w=\head{w}\tail{w}$. 
By $h(w)$ (head-action) we mean $a$
if $\head{w}=A$, and $b$ if $\head{w}=B$.
Subscripts $i,j$ range over $\{1,2,\dots,n\}$;
thus the rule $q_0 \goes{g} p_i$ stands for the $n$ rules
$q_0 \goes{g} p_1$, $q_0 \goes{g} p_2$, $\ldots$, $q_0 \goes{g} p_n$,
the rule
$p_i \goes{a_j} q_{0} I_j$, $i\neq j$, stands for $n\cdot(n{-}1)$
rules like $p_1 \goes{a_2} q_{0} I_2$, $p_8 \goes{a_5} q_{0} I_5$,
etc. (Rules with $(I^*)$ in (S1) are explained in the text.) 
\end{quote}

\caption{Rules from~\cite{DBLP:journals/jacm/JancarS08}, 
showing undecidability in the first-order 
case}\label{fig:rulesystem}
\end{figure}

Any position $(p\gamma,p\gamma)$ in the bisimulation game
is trivially winning for Defender. 
To avoid this ``equality-win'', when
starting from the position $(q_0 I_1\bot, q'_0 I_1\bot)$, 
Attacker obviously must not use the framed rule $q_0\goes{g}p_i$ 
(for any $i\in\{1,2,\dots,n\}$), nor $q'_0\goes{g}p_i$ which would
allow Defender to choose the framed rule to install equality. 
The frames just highlight
the use of Defender's forcing;
the rules are constructed so that Attacker must ensure
that neither him nor Defender ever uses a framed rule.

In the first round of the game, Attacker is thus forced
to use either $q_0 \goes{g} t$ ($g$ for ``generating'')
or $q_0 \goes{s} q_u$ ($s$ for ``switching'').
In the first case Defender uses
$q'_0 \goes{g} p_k$ for some (freely chosen) 
$k\in\{1,2,\dots,n\}$;
the current position becomes $(t I_1\bot, p_k I_1\bot)$.
Attacker is now forced to use $p_k \goes{a_k} q'_0 I_k$
or $t \goes{a_k} q_0 I_k$, since using an 
$a_j$-transition for $j\neq k$ allows Defender to install equality.
After Defender's response,
the current position is 
$(q_0 I_kI_1\bot, q'_0 I_kI_1\bot)$ where $k$ has been
chosen by Defender. 
We can thus see that the rules (G1) implement the generating phase. 
As long as Attacker chooses $g$, the play goes
through longer and longer pairs
\begin{equation*}%
(\, q_0\, I_{i_\ell}I_{i_{\ell-1}}\ldots I_{i_1}\bot\,,
\ q'_0\, I_{i_\ell}I_{i_{\ell-1}}\ldots I_{i_1}\bot\,)\,.
\end{equation*}
Since any infinite play is a win of Defender,
Attacker needs to enter the switching phase eventually, by using 
$q_0\goes{s}q_u$ from (S1).
The rules $q_0(I^*)I_i \goes{s} q_vw$, 
$q'_0(I^*)I_i \goes{s} q_vw$ constitute the only place where
$\varepsilon$-transitions enter the stage.
These rules stand for the 
following family of rules
given in (S1-$\tau$) in~\cite{DBLP:journals/jacm/JancarS08} 
(where $i$ ranges over $\{1,2,\dots,n\}$):
\begin{equation}\label{eq:familyepsilonrules}
q_0 \goes{s} z, \
q'_0 \goes{s} z, \
z I_i \goes{\varepsilon} z, \
z I_i \goes{\varepsilon} q_v w \
\ (\textnormal{for all suffixes } w
\textnormal{ of } v_i^R)\ .
\end{equation}
We note that the last rule 
$z I_i \goes{\varepsilon} q_v w$
could be made 
$\varepsilon$-popping by remembering $w$ 
in the control state but we prefer the given form for simplicity.
The $\varepsilon$-rules generate 
$\varepsilon$-transitions
in the respective \emph{fine-grained} LTS. 
Nevertheless 
we refer to the $\varepsilon$-free LTS where 
$s\goes{a}s'$ iff $s\goes{\varepsilon}\cdots \goes{\varepsilon}
s''\goes{a}s'''\goes{\varepsilon}\cdots \goes{\varepsilon}s'$
in the fine-grained LTS.

It is thus clear that Attacker is indeed forced to start 
the switching phase by choosing the rule $q_0\goes{s}q_u$
and performing the transition 
$q_0I_{i_\ell}I_{i_{\ell-1}}\ldots I_{i_1}\bot\goes{s}
q_uI_{i_\ell}I_{i_{\ell-1}}\ldots I_{i_1}\bot$.
Defender then chooses $m$ and $w$,
and the corresponding transition
$q'_0I_{i_\ell}I_{i_{\ell-1}}\ldots I_{i_1}\bot
\goes{s}
q_v wI_{i_m}I_{i_{m-1}}\ldots I_{i_1}\bot$
(where $w$ is a suffix of 
$(v_{i_{m+1}})^R$). 
The next current pair thus becomes
\begin{equation*}%
(\,q_u\, I_{i_\ell}I_{i_{\ell-1}}\ldots I_{i_1}\bot\,,
\ q_v\, w\,I_{i_m}I_{i_{m-1}}\ldots I_{i_1}\bot\,). 
\end{equation*}
Rules~(\ref{eq:familyepsilonrules}) also allow us to choose
$q'_0I_{i_\ell}I_{i_{\ell-1}}\ldots I_{i_1}\bot
\goes{s}
z I_{i_{m+1}}I_{i_{m-1}}\ldots I_{i_1}\bot$;
but once we understand the verification phase, 
we can easily check that this is of no help for Defender.
The verification phase is implemented by the rules (V1).
Defender can no longer threaten with installing equality but this is
not needed anymore; this phase is completely determined,
giving no real choice to any of the players.  
It is obvious that Defender  wins iff
\begin{equation*}%
(u_{i_{\ell}})^R(u_{i_{\ell-1}})^R\ldots (u_{i_{1}})^R=
w\,(v_{i_{m}})^R(v_{i_{m-1}})^R\ldots (v_{i_{1}})^R\ .
\end{equation*}
Since the described
bisimulation game closely mimicks our previous (intermediate) game,
it is easy to check that it also has Property~(\ref{eq:soliffwin}).

\section{Second-Order Pushdown Systems}\label{sec:undecsecondorder}

The ``first-order'' proof in Section~\ref{sec:undecepsilon}
(captured by the rules in Fig.~\ref{fig:rulesystem})
trivially shows the undecidability of bisimilarity for second-order
pushdown systems when 
$\varepsilon$-transitions are allowed.
When we explore the decidability question for
\emph{$\varepsilon$-free}
second-order pushdown systems then
it is natural to ask whether
we can implement the switching phase
(captured by (S1))
without using $\varepsilon$-rules, when
we have second-order \push and \pop at
our disposal.
So in terms of our intermediate game, we want to implement 
the switching from~(\ref{eq:switchstart}) 
to~(\ref{eq:switchend}). 
Without $\varepsilon$-transitions  we cannot implement erasing 
a prefix of $i_\ell i_{\ell-1}\cdots i_1$ 
(in the right-hand side string) in one move.
A natural idea is to shorten the right-hand side string 
step-by-step while Defender should 
decide when to finish. But it is not clear how 
to implement this in the ``first-order'' 
bisimulation game since Defender 
loses the possibility
of threatening with equality during such a step-by-step process. 
(S\'enizergues's decidability 
result~\cite{Senizergues:SIAM:05} shows that 
such an implementation is
indeed impossible in the first-order case.)

The idea (i.e., the crucial point in the undecidability proof 
in~\cite{DBLP:conf/fsttcs/BroadbentG12}) can be explained as follows.
When Attacker wants to switch at the position
$(q_0\, {i_\ell} {i_{\ell-1}}\dots {i_{1}},
\,q'_0\, {i_\ell} {i_{\ell-1}}\dots {i_{1}})$
then the stacks are doubled (using \push),
and the next position becomes
\begin{equation}\label{eq:doubled}
(r\, [{i_\ell} {i_{\ell-1}}\dots {i_{1}}][{i_\ell} {i_{\ell-1}}\dots {i_{1}}],
\,r'\, [{i_\ell} {i_{\ell-1}}\dots {i_{1}}][{i_\ell} {i_{\ell-1}}\dots
{i_{1}}])\ .
\end{equation}
Now the top stacks are being synchronously shortened, the play 
going through positions
$$(r\, [{i_{m+1}} {i_{m}}\dots {i_{1}}][{i_\ell} {i_{\ell-1}}\dots {i_{1}}],
\,r'\, [{i_{m+1}} {i_{m}}\dots {i_{1}}][{i_\ell} {i_{\ell-1}}\dots
{i_{1}}])$$
for decreasing $m$.
During this process
Defender can threaten with equality, so it is
possible to implement that it is Defender who decides when the process should
stop, forcing 
$\pop$ on the left-hand side (with entering $q_u$)
and choosing a suffix $w$ of $(v_{i_{m+1}})^R$ on the 
right-hand side; the reached position is then
\begin{equation} \label{eq:check}
(q_u\, [{i_\ell} {i_{\ell-1}}\dots {i_{1}}],
\ q_v\, [w\;{i_{m}}i_{m-1}\dots {i_{1}}][{i_\ell} {i_{\ell-1}}\dots
{i_{1}}]) \ . 
\end{equation}
The bottom stack on the right-hand side is now superfluous;
it only served for the previous threatening
with equality.
The verification phase is the same as previously
(with no choice for any player).
Implementing the described switching via the second-order rules 
is now a routine,
once we understand the Defender's forcing technique.
We just replace the rules (S1) in Fig.~\ref{fig:rulesystem}
with (S1-$2^{nd}$) in Fig.~\ref{fig:modifrules}
(where $i$ ranges over $\{1,2,\dots,n\}$).

\begin{figure}[ht]
\begin{tabular}{llll}
\vspace{1mm}
(S1-$2^{nd}$) rules:  & $q_0 \goes{s} (r, \push)$ 
 & & $q'_0 \goes{s} (r',\push)$\\ 
& $r \goes{c} q$ &  & 
$r' \goes{c} q'$, $r' \goes{c} q''$
\\
\vspace{1mm}
& \fbox{$r \goes{c} q'$, $r \goes{c} q''$}
\\ 
& $q I_i \goes{c_1} r$ && $q' I_i \goes{c_1} r'$ \\
\vspace{1mm}
& && \fbox{$q'' I_i \goes{c_1} r$} \\  
& $q \goes{c_2} p$ && $q'' \goes{c_2} p'$ \\
& && \fbox{$q' \goes{c_2} p$} \\ 
\vspace{1mm}
& $q\bot \goes{h} q$ \\   
& $p\goes{d}(q_u,\pop)$\\
& \fbox{$pI_i\goes{d}q_v w$} 
& & $p'I_i\goes{d}q_v w$ \hspace{3mm}
 (for each suffix $w$ of $(v_i)^R$)\\ \\
\end{tabular}
\caption{A replacement of (S1) to show undecidability
for $\varepsilon$-free second-order \pda}\label{fig:modifrules}
\end{figure}
\noindent
Now if Attacker chooses to switch (by action $s$) then
a position corresponding to (\ref{eq:doubled}) is reached
(where $i_j$ is replaced with $I_{i_j}$, and $\bot$ is added).
By Defender's forcing, Attacker must now use
the rule $r \goes{c} q$ and Defender decides whether
to enter the control state $q'$ (meaning that she wishes to erase
a further symbol $I_i$ 
from the top stacks, by the rules $qI_i \goes{c_1} r$
and $q'I_i \goes{c_1} r'$) or the control state $q''$ (meaning
that she wishes to enter the verification phase,
by the rules  $q \goes{c_2} p$
and $q'' \goes{c_2} p'$). In the next
round Attacker must follow these choices otherwise he will lose
(by the framed rules). 
The rule $q\bot \goes{h} q$ forces 
Defender to choose the second option (entering $q''$) 
before the top stacks are emptied (otherwise she loses).
Finally, once a position 
\begin{center}
$(p\, [I_{i_{m+1}} I_{i_{m}}\dots I_{i_{1}}\bot]
 [I_{i_\ell} I_{i_{\ell-1}}\dots I_{i_{1}}\bot],
\,p'\, [I_{i_{m+1}} I_{i_{m}}\dots I_{i_{1}}\bot]
 [I_{i_\ell} I_{i_{\ell-1}}\dots I_{i_{1}}\bot])$
\end{center}
is reached,
the last application of Defender's forcing
results in an analogue of~(\ref{eq:check}):
\begin{center}
$(q_u\, [I_{i_\ell} I_{i_{\ell-1}}\dots I_{i_{1}}\bot],
\,q_v\,[w\;I_{i_{m}} I_{i_{m-1}}\dots I_{i_{1}}\bot]
 [I_{i_\ell} I_{i_{\ell-1}}\dots I_{i_{1}}\bot])$\,.
\end{center}

\subsection{Normedness}

Bisimilarity problems like those we discuss here
are
often simpler when restricted to \emph{normed systems};
in our case, a \emph{state} $s$ in the LTS generated by  
a pushdown system is  \emph{normed} if from each
state that is reachable from $s$ we can reach 
a state where the stack is empty.
But restricting to the normed case does not affect the undecidability
here.
The states $q_0\, I_1\bot$, $q'_0\, I_1\bot$ in the system 
defined by
Fig.~\ref{fig:rulesystem} are normed, if 
we view the states $q_u\bot$, $q_v\bot$ as having the empty stack;
otherwise we can add
the rules
$q_u\bot \goes{e} q_u$,
$q_v\bot \goes{e} q_v$.
(In~\cite{DBLP:journals/jacm/JancarS08}, there are used
the rules $q_u\bot \goes{e} \varepsilon$,
$q_v\bot \goes{e} \varepsilon$ in the context 
of prefix-rewrite system definition.)

The authors of~\cite{DBLP:conf/fsttcs/BroadbentG12} 
are also interested in %
normedness for higher-order \pda as a natural extension of normedness for
first-order \pda.
We can note that after replacing (S1) in 
Fig.~\ref{fig:rulesystem}
with (S1-$2^{nd}$)
in Fig.~\ref{fig:modifrules}, the system is
not normed anymore 
(exemplified by the state
$q_v\,[\bot][I_{i_\ell}I_{i_{\ell-1}}...I_{i_1}\bot]$).
In~\cite{DBLP:conf/fsttcs/BroadbentG12} a triple
copy of the stack 
is used to handle the specific normedness definition there.
Another possibility is to start from
$(q_0\, [I_1\bot][\bot],q'_0\, [I_1\bot][\bot])$
and add a new control state $q_\textsc{pop}$ and
the rules $q_u \goes{f}q_\textsc{pop}$ and
$x \goes{f}(q_\textsc{pop},\pop)$ for all control states $x$
where $x \not=q_u$, 
including the rule $q_\textsc{pop}\goes{f}(q_\textsc{pop},\pop)$.

\subsection*{Additional comments}

As already mentioned, the undecidability result 
for $\varepsilon$-free 
second-order pushdown systems 
in~\cite{DBLP:conf/fsttcs/BroadbentG12}
clarifies the (un)decidability border in another direction 
than the undecidability result 
for first-order pushdown systems with $\varepsilon$-transitions
in~\cite{DBLP:journals/jacm/JancarS08}.
The border can be surely explored further. 
For example it seems that we cannot avoid 
using several control states in the above undecidability proofs
(though we can surely decrease their number 
by extending the stack alphabet).
Hence (normed) second-order simple grammars, 
studied in~\cite{DBLP:conf/concur/Stirling06}, 
are a possible target for exploring.
\bibliographystyle{abbrv}
\bibliography{concur}

\end{document}